\documentclass[preprint,prl,showpacs,showkeys]{revtex4}
\usepackage{graphicx}
\usepackage{dcolumn}
\usepackage{amsmath}
\usepackage{bm}
\newcommand{\be}{\begin{equation}}
\newcommand{\ee}{\end{equation}}
\newcommand{\ba}{\begin{eqnarray}}
\newcommand{\ea}{\end{eqnarray}}
\begin{document}
\title{Anomalous phase behavior in a model fluid \\
with only one type of local structure}
\author{Santi Prestipino$^1$~\cite{aff1}, Franz Saija$^2$~\cite{aff2},
and Gianpietro Malescio$^1$~\cite{aff3}}
\affiliation{
$^1$ Universit\`a degli Studi di Messina, Dipartimento di Fisica,
Contrada Papardo, 98166 Messina, Italy \\
$^2$ CNR-Istituto per i Processi Chimico-Fisici, Viale Ferdinando Stagno
d'Alcontres 37, 98158 Messina, Italy}
\date{\today}
\begin{abstract}

We present evidence that the concurrent existence of two populations of
particles with different effective diameters is not a prerequisite
for the occurrence of anomalous phase behaviors in systems of particles
interacting through spherically-symmetric unbounded potentials.
Our results show that an extremely weak softening of the interparticle
repulsion, yielding a single nearest-neighbor separation, is able
to originate a wide spectrum of unconventional features including
reentrant melting, solid polymorphism, as well as thermodynamic, dynamic,
and structural anomalies.
These findings extend the possibility of anomalous phase behavior to a
class of systems much broader than currently assumed.

\end{abstract}
\maketitle

\section{I. Introduction}

The phase behavior of one-component substances is termed ``anomalous''
whenever it differs from that of prototypical (i.e., argon-like) liquids.
Anomalous phase behavior includes polymorphism in the liquid and solid
phases, reentrant melting (i.e., melting by compression at constant
temperature), and a host of other thermodynamic, dynamic, and structural
anomalies. An important class of systems displaying such features is that
of network-forming fluids, i.e., fluids that form orientation-specific,
intermolecular bonds which are strong relative to London forces.
The most well-known of these substances is water~\cite{Debenedetti0,Mishima},
while other cases are silicon, phosphorous, and
silica~\cite{Sastry,Katayama,Meade}.
Anomalous behaviors have been observed also in systems where effective
interatomic forces can reasonably be assumed non-directional. For example,
alkali and alkali-earth metals at high pressures exhibit one or more regions
of reentrant melting~\cite{Young}. Liquid or amorphous polymorphism have been
found in other systems with non-directional interactions, such as molten
Al$_2$O$_3$-Y$_2$O$_3$~\cite{Aasland}, triphenyl phosphite~\cite{Tanaka},
and some metallic glasses~\cite{Sheng}. 

The quest for understanding the basic mechanisms of anomalous phase behaviors
has given impulse to the study of simple isotropic potentials that are able
to display the same behaviors.
These systems can be appropriate models for the generic
thermodynamic behavior of pure metals, metallic mixtures, electrolytes, and
colloids. It has been found that unusual behaviors may arise in systems
of spherical particles (simple fluids) where the unbounded repulsive core
is ``softened'' through the addition of a finite repulsion at intermediate
distances, so as to generate two distinct length scales in the system:
a ``hard'' one, related to the inner core, and a ``soft'' one, associated
with the soft, penetrable, component of the
repulsion~\cite{Hemmer,SadrLahijany,Jagla,Franzese,Watzlawek,Yan,Gibson,Malescio1,Fomin,Pauschenwein,Malescio2,Deoliveira,Saija1,Lascaris}. 
Due to this feature, such core-softened (CS) fluids are characterized by two
competing, expanded and compact, local arrangements of particles.
This property mimics
the behavior of the more complex network-forming fluids, where loose and
compact local structures arise from the continuous formation and disruption
of the dynamic network originated by orientational bonds.
So far, the existence of two competing local structures has been regarded
as essential for the occurrence of anomalous phase behaviors in simple
fluids. In this work we challenge this belief and provide evidence that
these behaviors may also occur in systems with a single local particle
arrangement, i.e., under much weaker conditions than admitted so far.

\section{II. Theory and simulation}

For a spherically-symmetric unbounded potential $u(r)$ the analytic condition
for core softening~\cite{Debenedetti1} is that, in a range of interparticle
distances $r$ within the core, the product $rf(r)$, with
$f(r)=-{\rm d}u/{\rm d}r$, decreases as $r$ gets smaller. Systems satisfying
this core-softening condition are characterized by two different repulsive
length scales and, for pressures where the two scales are both effective,
loose and compact local arrangements compete with each other and the
fluid behaves like a mixture of two particle species with different radii
(``two-state'' fluid~\cite{Rapoport,McMillan,Debenedetti2}).
A typical property of such fluids is the existence of a reentrant melting
line, a feature that is strictly related to other anomalous
behaviors~\cite{McMillan}. We note that the core-softening condition is not
applicable to lattice models with a softened two-body repulsion, which can
nonetheless show a reentrant-melting behavior~\cite{Hoye,Almarza}.

In order to assess whether or not the existence of two distinct local
structures is a necessary requirement for reentrant melting, we ask
the following question: how much soft should repulsion be in 
order that reentrant melting may occur in simple fluids?
To attempt an answer, we have developed a convenient tool for the analysis
of trends in the melting-line topology, based on linear-elasticity
theory~\cite{Barron}. We focus on the face-centered cubic (fcc) and the
body-centered cubic (bcc) phases, i.e., those relevant for purely-repulsive
potentials at low to moderate pressure. According to elasticity theory,
the enthalpy of a cubic-symmetry solid under homogeneous strain is
\be
H=H_0+V_0\left\{\frac{1}{2}\lambda_1(e_{xx}^2+e_{yy}^2+e_{zz}^2)
+\lambda_2(e_{xx}e_{yy}+e_{xx}e_{zz}+e_{yy}e_{zz})
+2\lambda_3(e_{xy}^2+e_{xz}^2+e_{yz}^2)\right\}\,,
\label{eq1}
\ee
where $H_0$ is the enthalpy of the undeformed crystal with volume $V_0$,
$e_{\alpha\beta}$ are the strain-tensor components, and the $\lambda$'s
are Lam\'e coefficients. At zero temperature, these coefficients are
given in terms of the interparticle potential~\cite{Rechtsman}, being
ultimately a function of pressure.
Then, we employ Eq.\,(\ref{eq1}) in a field theory
where the field variables are the components of the displacement vector
${\bf d}({\bf R})$ at each lattice site.
In practice, any given realization of the displacement field is assigned
the weight $\exp\{-H/(k_BT)\}$, with the $\lambda$'s fixed at their
$T=0$ values. In Fourier space, this eventually gives the mean square
displacement as a Gaussian integral. We find
\be
\left\langle d^2\right\rangle=
k_BT\int_{\rm DS}\frac{{\rm d}^3q}{(2\pi)^3}{\rm Tr}[M^{-1}({\bf q})]\,,
\label{eq2}
\ee
where the sum in (\ref{eq2}) is over the Debye sphere (DS) and
$M^{-1}({\bf q})$ is the inverse of
\be
M_{\alpha\beta}({\bf q})=[\lambda_3q^2+(\lambda_1-\lambda_2-2\lambda_3)
q_\alpha^2]\delta_{\alpha\beta}+(\lambda_2+\lambda_3)q_\alpha q_\beta\,.
\label{eq3}
\ee
Derivation of Eq.(\ref{eq2}) assumes that the quadratic form in
(\ref{eq1}) is positive definite, meaning that the stability conditions
$\lambda_1+2\lambda_2>0,\lambda_1-\lambda_2>0$, and $\lambda_3>0$
must hold. At those pressures where any of these inequalities fails to be
satisfied, the solid is mechanically unstable.

We obtain a closed-form expression for the melting temperature $T_m(P)$
by using Lindemann's criterion, i.e., by equating the r.h.s. of
Eq.(\ref{eq2}) for $T=T_m$ to $(\delta_L r_{\rm NN})^2$, where
$\delta_L\approx 0.15$ for a fcc solid and $\delta_L\approx 0.18$ for
a bcc solid~\cite{Saija2}, and $r_{\rm NN}$ is the nearest-neighbor (NN)
distance. We have checked in known cases (Lennard-Jones, exp-6, and
Gaussian potentials) that the present criterion correctly predicts the
overall trend of the fcc and bcc melting lines, although $T_m$ is
systematically overestimated, which is not surprising in view of
the assumption of temperature-independent $\lambda$'s.

\begin{figure}
\includegraphics[width=10cm]{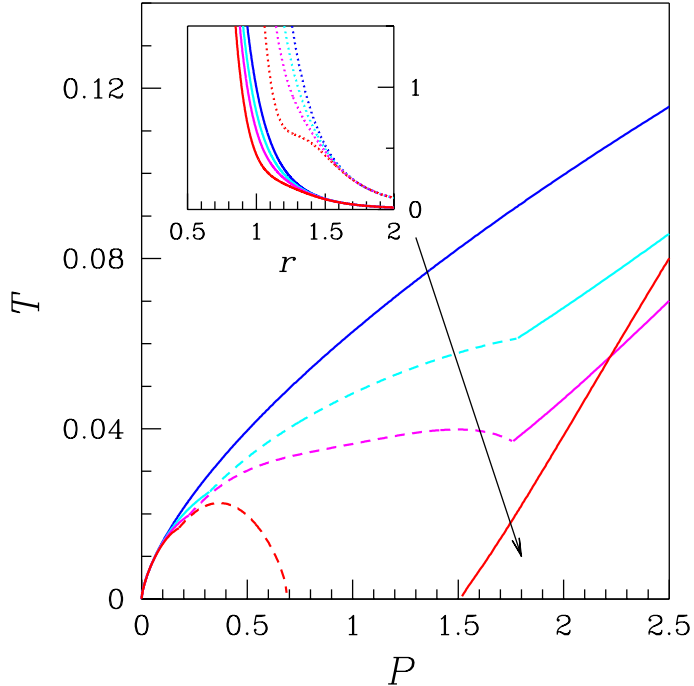}
\caption{(Color online). 
Theoretical melting lines (see text) for $u_{\rm MIP}(r)$.
Pressure $P$ and temperature $T$ are in reduced, $\epsilon/\sigma^3$
and $\epsilon/k_B$, units.
Melting lines are obtained by taking, for each $P$, the higher $T_m$ value
between fcc (solid line) and bcc (dashed line).
Four cases are shown: $A=0$ (blue), $-0.2$ (cyan), $-0.35$
(magenta), $-0.55$ (red). Repulsion softness grows in the arrow direction.
Cusps at estimated fluid-fcc-bcc triple points are artefacts of the
approximation. Upon increasing the repulsion softness further and
further, a stability gap eventually opens at $T=0$ for both fcc and bcc
solids, which will be filled by one or more low-coordinated crystals. 
In the inset, the four potentials are plotted as continuous lines,
with the respective $rf(r)$ also shown as dotted lines.
}
\label{fig1}
\end{figure}

\begin{figure}
\includegraphics[width=10cm]{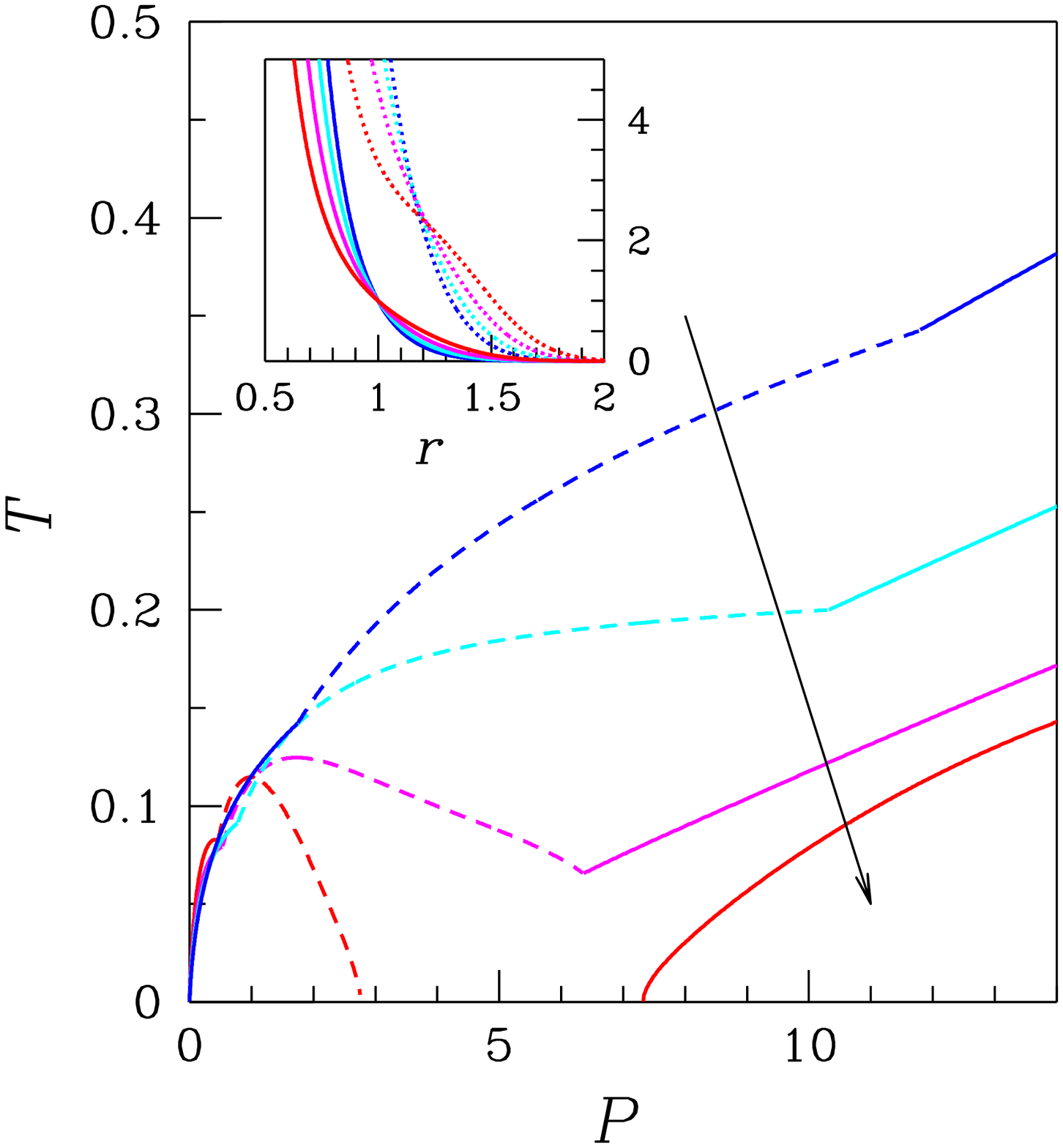}
\caption{(Color online). 
Theoretical melting lines (see text) for $u_{\rm YK}(r)$.
Four cases are shown: $a=6.9$ (blue), $5.7$ (cyan), $4.5$ (magenta),
$3.3$ (red). Repulsion softness grows in the arrow direction.
Upon increasing the repulsion softness further and further,
a stability gap eventually opens at $T=0$ for both fcc and bcc
solids, which will be filled by one or more low-coordinated crystals. 
In the inset, the four potentials are plotted as continuous lines,
with the respective $rf(r)$ also shown as dotted lines.
}
\label{fig2}
\end{figure}

We now apply the above outlined melting criterion to two different types of
repulsive interactions with tunable softness. 
We first consider the modified inverse-power (MIP) potential $u_{\rm MIP}(r)=
u_0(r) [1+A\exp\left\{-10(1-r/\sigma)^2\right\}]$ with
$u_0(r)=\epsilon(r/\sigma)^{-6}$, $\epsilon$ and $\sigma$ being energy and
length units. As $|A|$ increases, core repulsion becomes softer and the
core-softening condition is satisfied for $A\lesssim -0.60$ and
$A\gtrsim 28$. Our criterion predicts reentrant melting for $A\lesssim -0.35$
(Fig.\,1) and $A\gtrsim 1$ (not shown). 
We perform similar calculations for a different interaction law,
$u_{\rm YK}(r)=\epsilon\exp\left\{
a(1-r/\sigma)-6(1-r/\sigma)^2\ln(r/\sigma)\right\}$,
introduced long ago by Yoshida and Kamakura (YK)~\cite{Yoshida}.
This repulsion becomes softer with decreasing $a$
and satisfies the core-softening condition for $a\lesssim 2.3$. The
theoretical melting line shows a reentrant portion for $a\lesssim 5.5$
(see Fig.\,2).

The above results intriguingly suggest that reentrant melting (and
presumably also other anomalous behaviors) can occur for continuous
potentials that {\it do not} satisfy the core-softening condition.
To assess this possibility, we investigate through numerical simulation
the equilibrium properties of a system of particles interacting through
$u_{\rm YK}(r)$ for $a=3.3$. This potential does not satisfy the core-softening
condition and monotonously increases, for decreasing $r$, together with
its first and second derivatives (see Fig.\,3 inset).
To carry on the simulation study, we assume that the same crystals that are
stable at $T=0$ also give the underlying lattice structure for the stable
solid phases at $T>0$. Through a total-energy calculation similar to that
in \cite{Prestipino}, we find that at $T=0$ the sequence of stable crystals
for increasing pressures is (see \cite{Prestipino} for acronyms)
\be
{\rm fcc}
\stackrel{0.66}{\longrightarrow}{\rm bcc}
\stackrel{2.51}{\longrightarrow}\beta{\rm Sn}
\stackrel{3.50}{\longrightarrow}{\rm sh}
\stackrel{3.86}{\longrightarrow}{\rm sc}
\stackrel{5.07}{\longrightarrow}\beta{\rm Sn}
\stackrel{7.22}{\longrightarrow}{\rm cI16}
\stackrel{8.33}{\longrightarrow}{\rm fcc}
\stackrel{16.7}{\longrightarrow}{\rm hcp}
\stackrel{64.8}{\longrightarrow}{\rm fcc}
\stackrel{282}{\longrightarrow}{\rm hcp}
\stackrel{518}{\longrightarrow}{\rm fcc}\,,
\label{eq4}
\ee
where the numbers are the transition pressures in units of
$\epsilon/\sigma^3$. We note that the internal parameter of the stable
cI16 crystal varies from 0.043 to 0.047, incidentally close to the cI16
phases of Li~\cite{Hanfland} and Na~\cite{Mcmahon}.

\begin{figure}
\includegraphics[width=10cm]{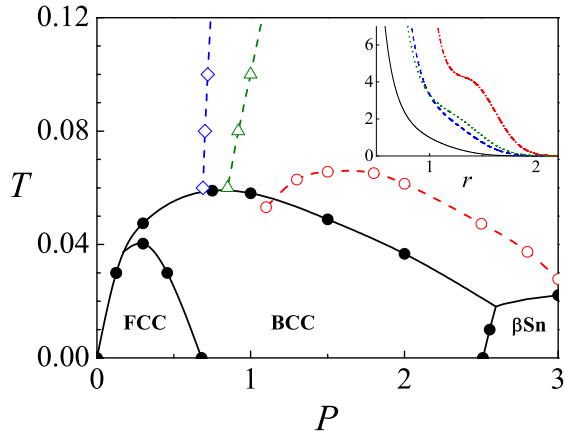}
\caption{(Color online). Phase diagram of the YK potential for $a=3.3$
($P$ and $T$ are in reduced units). Full dots are two-phase coexistence
points. Open dots are points of density maximum in the fluid phase.
Diamonds and triangles denote points of $-s_2$ maxima and $D$ minima
respectively ($D$ being the self-diffusion coefficient), giving the
left boundary of the regions of structural and diffusion anomaly
(the right boundaries, defined by $-s_2$ minima and $D$ maxima,
are out of the $P$ range shown). Numerical errors are smaller than the
symbols size. Within the pressure range
$2.5\div 3$, the aspect ratio of the $\beta$Sn phase is $c/a\simeq 0.60$.
Inset: $u_{\rm YK}(r)$ for $a=3.3$ (black solid line), $f(r)=-u'_{\rm YK}(r)$
(blue dashed line), $rf(r)$ (green dotted line), $u''_{\rm YK}(r)$
(red dot-dashed line).}
\label{fig3}
\end{figure}

We have performed extensive Monte Carlo (MC) and molecular-dynamics
simulations, using samples of $500$ to $800$ particles (simulations
with 2048 particles lead to practically the same results).
Coexistence lines are computed by exact free-energy methods (details
of the MC simulation are as in \cite{Saija1}). The phase diagram
for $P<3$ is plotted in Fig.\,3. The bcc melting temperature shows a
maximum at $P\simeq 0.75$, hence at not too low temperatures
the bcc solid melts upon compression into a denser fluid. At still lower
temperatures, the bcc phase transforms into a $\beta$Sn phase.
In the reentrant-fluid region a density anomaly occurs, i.e., the number
density decreases upon cooling at constant pressure. This region is
bounded from above by the temperature of the maximum density line
(see Fig.\,3). In Fig.\,3, we also show points of diffusivity minimum and
of $-s_2$ maximum, $s_2$ being the two-body entropy~\cite{Esposito}.
Similarly to water, the region of structural anomaly encompasses that of
anomalous diffusion, which in turn encloses the region of density anomaly.
A compendium of the anomalous behavior of the present system
is shown in Fig.\,4.

\begin{figure}
\includegraphics[width=10cm]{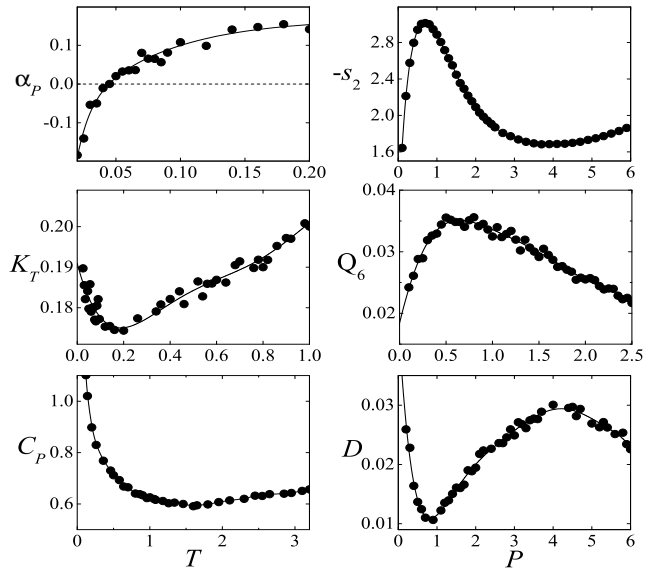}
\caption{Thermodynamic and structural quantities for the YK fluid
with $a=3.3$.
Left column: thermal expansion coefficient $\alpha_P$ (units of
$k_B/\epsilon$), isothermal compressibility $K_T$ (units of
$\sigma^3/\epsilon$), and constant-pressure specific heat $C_P$
(units of $k_B$) as a function of $T$ along the isobar $P=2.5$. 
For conventional liquids, $\alpha_P$, $K_T$, and $C_P$ monotonically
increase with $T$ and $\alpha_P>0$. Right column:
translational order parameter $-s_2$ (units of $k_B$), bond-order parameter
$Q_6$~\cite{Steinhardt}, and self-diffusion coefficient $D$ (units
of $\sigma(\epsilon/m)^{1/2}$, where $m$ is particle mass) as a
function of $P$ along the isotherm $T=0.06$.
For conventional liquids, $-s_2$ and $Q_6$ increase with $P$
while $D$ decreases monotonically.}
\label{fig4}
\end{figure}

\section{III. Discussion}

The anomalies exhibited by the YK fluid for $a=3.3$ are quite similar
to those of CS potentials. However, the much weaker softness of the former 
leads to peculiar local arrangement of particles in the fluid phase,
as evidenced in the radial distribution function $g(r)$.
As $P$ increases at constant temperature, the NN peak of $g(r)$ moves
gradually toward small $r$ (Fig.\,5, top panel). Meanwhile its height
first grows, due to increasing proximity with the bcc solid, and then goes 
down in the pressure range where reentrant melting occurs. As $P$ increases
further, the NN peak of $g(r)$ grows again while its position changes less
and less sensibly due to the steep small-$r$ repulsion. 
At low pressures, the behavior of $g(r)$ is similar to that of the
Gaussian-core model~\cite{Likos} (GCM), a fluid with a bounded interaction
potential that also exhibits a bunch of waterlike anomalies. However,
due to the absence of a true particle core, the behavior is completely
different at high pressures where the GCM more and more resembles an
``infinite-density ideal gas'' with a radial distribution function $g(r)=1$.
To our knowledge, the $g(r)$ behavior of the system investigated here has
never been reported for other systems of truly impenetrable particles.

To make clear the difference between the interaction studied here
and the CS potentials investigated so far, we report the radial
distribution function of a CS potential (Fig.\,5, bottom panel). 
In this case, the heights of the first two peaks of $g(r)$, associated
respectively with the hard and soft length scales, change in opposite
directions on increasing pressure, signalling that the ``hard'' NN distance 
becomes more and more populated at the expenses of the ``soft'' distance,
while the position of the two peaks remains essentially unaltered. 
This behavior is a clear evidence of the simultaneous existence of two 
populations of particles having distinct effective diameters, a
phenomenon that does not occur in the case analyzed here.

\begin{figure}
\includegraphics[width=10cm]{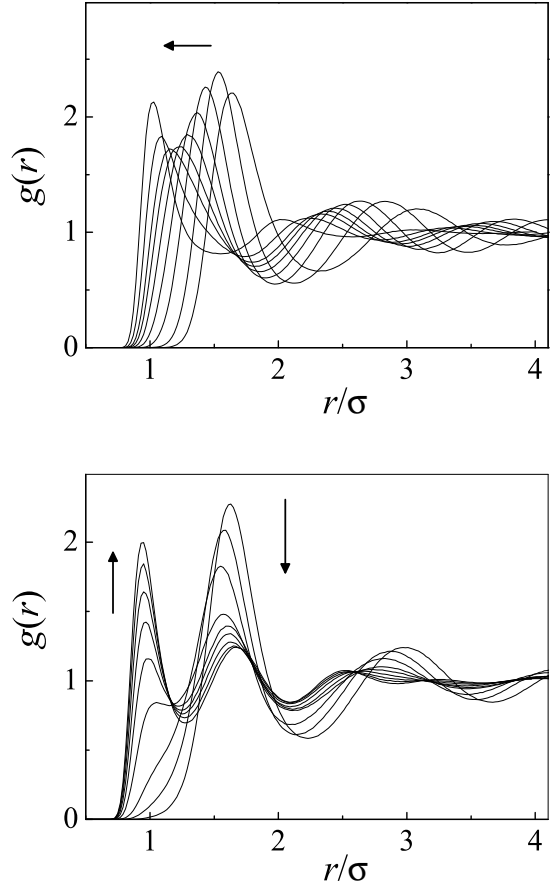}
\caption{Top panel: pair distribution function $g(r)$ of the YK potential
for $a=3.3$. Bottom panel: $g(r)$ of the YK potential for $a=2.1$ (not far
from the core-softening threshold $a=2.3$). For $a=3.3$, lines correspond
to $T=0.06$ and $P=0.1,0.5,1,1.5,2,2.5,3,4,6$. For $a=2.1$, curves refer to
$T=0.07$ and $P=0.5,0.7,1,1.6,2.3,3,4.1,5.2,6.4$. The arrows mark the
direction of pressure increase.} 
\label{fig5}
\end{figure}

\begin{figure}
\includegraphics[width=10cm]{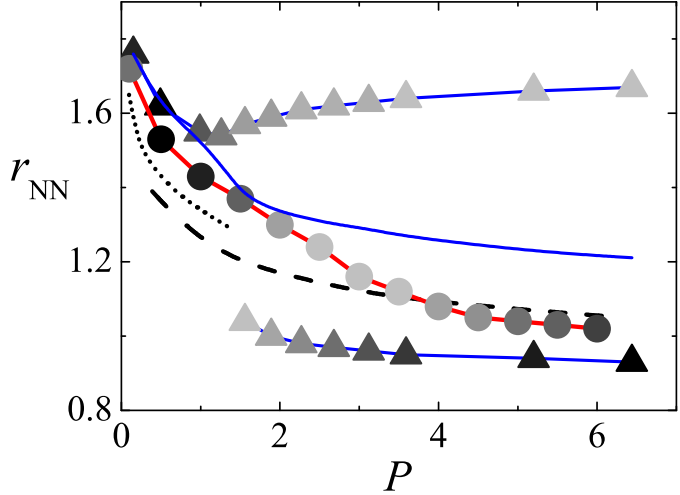}
\caption{Position $r_{\rm NN}$ of the NN peak of
$g(r)$ in units of $\sigma$ as a function of $P$ at constant $T$ for:
$u_{\rm YK}(r)$, $a=2.1$ and $T=0.07$ (blue solid line and triangles);
$u_{\rm YK}(r)$, $a=3.3$ and $T=0.06$ (red solid line and full dots);
$u_0(r)$, $T=0.06$ (black dotted line, stopping near the melting point);
$u_0(r)$, $T=1$ (black dashed line). The gray scale is proportional to the
height of the $g(r)$ peak.
The blue solid line without symbols represents a weighted average of the
two $r_{\rm NN}$ branches, with weights proportional to the respective $g(r)$
peak heights, for the case $u_{\rm YK}(r)$ with $a=2.1$.}
\label{fig6}
\end{figure}

To gain further insight we analyze the $P$ dependence of the NN peak position
$r_{\rm NN}$ for the system under study, as compared with a CS potential 
($u_{\rm YK}(r)$ with $a=2.1$) and $u_0(r)$ (see Fig.\,6). 
For the CS potential, $r_{\rm NN}(P)$ consists of two branches, i.e., in a
range of pressures two distinct populations of particles with different
effective radii coexist, which legitimates the two-state-fluid picture.
On the contrary, for the present system $r_{\rm NN}(P)$
has only one branch (similarly to $u_0(r)$ where, however, the $g(r)$ peaks
grow monotonously with $P$), meaning that all particles have the same
effective radius. While the $r_{\rm NN}(P)$ curve for $u_0(r)$ is upward
concave everywhere, in the present case it shows a portion with downward
concavity, roughly corresponding to the reentrant-melting pressure range.
The same feature is displayed by the weighted average of the two $r_{\rm NN}$
branches for the CS potential (the downward-concave region is here shifted,
as well as the reentrant portion of the melting line, towards smaller
pressures).

Thus, in the case considered here there exists only one type of local
structure with an effective length scale that, however, shrinks with pressure
in such a way as to capture the $P$-dependence of the mean length scale for
the CS case. In particular, the shrinking rate shows a local maximum in the
reentrant-melting pressure range, which makes the length scale more loosely
defined with respect to the adjacent regions. As shown by our results,
this is sufficient to give origin to anomalous phase behavior.

\section{IV. Concluding remarks}

Competition between two different local particle arrangements, arising
from either directional or core-softened isotropic forces, has so far
deemed to be responsible for anomalous thermodynamic behavior. In this
article, we have shown that the same behaviors may occur also for
weakly-softened (WS) potentials, i.e., simple fluids characterized by a
repulsion that is only marginally softened and yields a single structure
at a local level. Our results overturn the two-state-fluid picture derived
from network-forming fluids and simplify the minimal scenario for the
occurrence of anomalous behaviors, thus considerably widening the class of
interactions that give rise to waterlike anomalies.
WS potentials can be relevant in the realm of soft matter, where engineering
interparticle forces is possible, and also for ``hard'' matter under extreme
conditions. In the latter case, pressure typically induces successive
rearrangements of the crystal structure in order to minimize the electronic
energy. This is reflected at higher $T$ in reentrant melting and (though
more rarely reported~\cite{Thurn}) waterlike anomalies. The effect of
pressure on the atomic structure may be dramatic, like e.g. in Cs where
a 6s-5d band crossing causes a sudden drop in the effective radius of the
atom~\cite{Sternheimer} (``orbital collapse''), or less sharp as can be
expected for lighter and smaller chemical elements. While two-scale CS
potentials offer a simple model for systems undergoing orbital collapse,
one-scale WS potentials may be appropriate to describe those elements
where the atomic radius shrinks more gradually under pressure.


\begin{thebibliography}{99}
\bibitem[*]{aff1} E-mail: {\tt sprestipino@unime.it}

\bibitem[\dag]{aff2} Corresponding author. E-mail: {\tt saija@me.cnr.it}

\bibitem[\ddag]{aff3} E-mail: {\tt malescio@unime.it}

\bibitem{Debenedetti0} P. G. Debenedetti {\em Metastable Liquids}
(Princeton University Press, Princeton, 1996).

\bibitem{Mishima} O. Mishima and H. E. Stanley, Nature {\bf 396}, 329 (1998).

\bibitem{Sastry} S. Sastry and C. A. Angell, Nature Mater. {\bf 2}, 739 (2003).

\bibitem{Katayama} Y. Katayama {\it et al.}, Nature {\bf 403}, 170 (2000). 

\bibitem{Meade} C. Meade {\it et al.}, Phys. Rev. Lett. {\bf 69}, 1387 (1992).

\bibitem{Young}  D. A. Young, {\em Phase Diagrams of the Elements}
(University of California, Berkeley, 1991).

\bibitem{Aasland}  S. Aasland and P. F. McMillan, Nature {\bf 369}, 633 (1994).

\bibitem{Tanaka}  H. Tanaka {\it et al.},
Phys. Rev. Lett. {\bf 92}, 025701 (2004).

\bibitem{Sheng}  H. W. Sheng {\it et al.},
Nature Materials {\bf 6}, 192 (2007).

\bibitem{Hemmer}  P. C. Hemmer and G. Stell,
Phys. Rev. Lett. {\bf 24}, 1284 (1970).

\bibitem{SadrLahijany}  M. R. Sadr-Lahijany {\it et al.},
Phys. Rev. Lett. {\bf 81}, 4895 (1998).

\bibitem{Jagla}  E. A. Jagla, Phys. Rev. E {\bf 58}, 1478 (1998).

\bibitem{Watzlawek}  M. Watzlawek {\it et al.},
Phys. Rev. Lett. {\bf 82}, 5289 (1999).

\bibitem{Franzese}  G. Franzese {\it et al.},
Nature {\bf 409}, 692 (2001).

\bibitem{Yan}  Z. Yan {\it et al.},
Phys. Rev. Lett. {\bf 95}, 130604 (2005).

\bibitem{Gibson}  H. M. Gibson and N. B. Wilding,
Phys. Rev. E {\bf 73}, 061507 (2006).

\bibitem{Malescio1} G. Malescio,
J. Phys.: Condensed Matter {\bf 19}, 073101 (2007)

\bibitem{Fomin}  D. Yu. Fomin {\it et al.},
J. Chem. Phys. {\bf 129}, 064512 (2008).

\bibitem{Pauschenwein}  G. J. Pauschenwein and G. Kahl,
Soft Matter {\bf 4}, 1396 (2008).

\bibitem{Malescio2}  G. Malescio {\it et al.},
J. Chem. Phys. {\bf 129}, 241101 (2008).

\bibitem{Deoliveira}  A. B. de Oliveira {\it et al.},
Europhys. Lett. {\bf 85}, 36001 (2009).

\bibitem{Saija1}  F. Saija {\it et al.},
Phys. Rev. E {\bf 80}, 031502 (2009).

\bibitem{Lascaris}  E. Lascaris {\it et al.}, Phys. Rev. E {\bf 81},
031201 (2010).

\bibitem{Debenedetti1}  P. G. Debenedetti {\it et al.},
J. Phys. Chem. {\bf 95}, 4540 (1991).

\bibitem{Rapoport}  E. Rapoport, J. Chem. Phys. {\bf 46}, 2891 (1967).

\bibitem{McMillan}  P. F. McMillan, J. Mater. Chem. {\bf 14}, 1506 (2004).

\bibitem{Debenedetti2}  P. G. Debenedetti and H. E. Stanley,
Physics Today {\bf 56}, 40 (2003).

\bibitem{Hoye}  J. S. H\o ye {\it et al.}, Mol. Phys. {\bf 107}, 321 (2009).

\bibitem{Almarza}  N. G. Almarza {\it et al.}. J. Chem. Phys. {\bf 131},
124506 (2009).

\bibitem{Barron}  See e.g. T. H. K. Barron and M. L. Klein,
Proc. Phys. Soc. {\bf 85}, 523 (1965).

\bibitem{Rechtsman}  M. C. Rechtsman {\it et al.},
Phys. Rev. Lett. {\bf 101}, 085501 (2008).

\bibitem{Saija2}  F. Saija {\it et al.},
J. Chem. Phys. {\bf 124}, 244504 (2006).

\bibitem{Yoshida}  T. Yoshida and S. Kamakura, 
Prog. Theor. Phys. {\bf 52}, 822 (1974).

\bibitem{Prestipino}  S. Prestipino {\it et al.},
Soft Matter {\bf 5}, 2795 (2009).

\bibitem{Hanfland}  M. Hanfland {\it et al.},
Nature {\bf 408}, 174 (2000).

\bibitem{Mcmahon}  M. I. McMahon {\it et al.},
Proc. Natl. Acad. Sci. USA {\bf 104}, 17297 (2007).

\bibitem{Esposito}  See e.g. R. Esposito {\it et al.},
Phys. Rev. E {\bf 73}, 040502(R) (2006).

\bibitem{Steinhardt}  P. J. Steinhardt {\it et al.},
Phys. Rev. B {\bf 28}, 784 (1983). 

\bibitem{Likos}  C. N. Likos, Phys. Rep. {\bf 348}, 267 (2001).

\bibitem{Thurn}  H. Thurn and J. Ruska,
J. Non-Cryst. Solids {\bf 22}, 331 (1976).

\bibitem{Sternheimer}  R. Sternheimer, Phys. Rev. {\bf 78}, 235 (1950).

\end{thebibliography}
\end{document}